\documentclass[twocolumn, preprintnumbers,prb,aps,amssymb,showpacs]{revtex4}

\usepackage{boxedminipage}
\usepackage{lscape}
\usepackage{float}
\usepackage[ansinew]{inputenc}
\usepackage{epsfig}
\usepackage{subfigure}
\usepackage{graphicx}
\usepackage{amsmath}
\usepackage{amssymb}
\usepackage{amsthm,url}
\usepackage[francais]{babel}
\usepackage{color}
\usepackage{natbib}
\usepackage{dcolumn}
\usepackage{wasysym}
\usepackage{bm}

\usepackage{url}
\usepackage[pdftex,plainpages=false,colorlinks=true,linkcolor=blue, citecolor=blue, urlcolor=blue]{hyperref}
\usepackage[figure]{hypcap}
\usepackage{graphicx}

\begin{document}

\title{Mott p-n Junctions in layered materials}
\author{M. Charlebois$^{1}$, S.R. Hassan$^{2}$, R. Karan$^{2}$, D. Sénéchal$^{1}$, A.-M. S. Tremblay$^{1,3}$}
\affiliation{$^1$ Département de Physique and RQMP, Université de Sherbrooke, Sherbrooke,
QC, Canada\\
$^{2}$The Institute of Mathematical Sciences C.I.T. Campus, Chennai 600 113, India\\
$^{3}$Canadian Institute for Advanced Research, Toronto, Ontario, Canada.}
\date{\today}
\keywords{heterostructure Mott Hubbard CDMFT p-n junction}
\pacs{74.78.Fk, 71.30.+h, 71.10.Fd}

\begin{abstract}
The p-n junction has provided the basis for the semiconductor-device industry. Investigations of p-n junctions based on Mott insulators is still in its infancy. Layered Mott insulators, such as the cuprates or other transition metal-oxides, present a special challenge since strong in-plane correlations are important. Here we model the planes carefully using plaquette Cellular Dynamical Mean Field Theory with an exact diagonalization solver. The energy associated with inter-plane hopping is neglected compared with the long-range Coulomb interaction that we treat in the Hartree-Fock approximation. Within this new approach, ``Dynamical Layer Theory'', the charge redistribution is obtained at the final step from minimization of a function of the layer fillings. A simple analytical description of the solution, in the spirit of Thomas-Fermi theory, reproduces quite accurately the numerical results. Various interesting charge reconstructions can be obtained by varying the Fermi energy differences between both sides of the junction. One can even obtain quasi-two dimensional charge carriers at the interface, in the middle of a Mott insulating layer. The density of states as a function of position does not follow the simple band bending picture of semiconductors. 
\end{abstract}

\maketitle



\section{Introduction}

Strong electronic correlations in transition-metal oxides manifest themselves in many spectacular ways, including high-temperature superconductors, Mott insulators, exotic magnetic phases and colossal magnetoresistance materials.~\cite{dagotto_complexity_2005}. Although these phases are present in the bulk, we expect to obtain even more complex and interesting states of matter if we harvest the properties of correlated electron with the help of heterostructures~\cite{dagotto_when_2007,ramirez_oxide_2007,heber_materials_2009,mannhart_oxide_2010,hwang_emergent_2012}. A decade ago, Ohtomo et al. \cite{ohtomo_artificial_2002} demonstrated the technical capability to create atomically sharp interfaces between transition metal oxides, launching a revolution in the field of oxide heterostructures.

The field is presently dominated by the well known LaAlO$_3$/SrTiO$_3$ interface, which gained much popularity because it harbours a two-dimensional electron gas~\cite{ohtomo_high-mobility_2004}. In the bulk, both oxides are insulators and non-magnetic materials. However, when we consider the interface, it can become superconducting \cite{reyren_superconducting_2007}, ferromagnetic~\cite{brinkman_magnetic_2007} and both phases can even coexist~\cite{li_coexistence_2011}. These effects can all be explained by charge transfer. It is thus possible with an external applied electric field to explore the phase diagram~\cite{ahn_electric_2003, caviglia_electric_2008,ueno_electric-field-induced_2008} and create exotic interface effects~\cite{caviglia_tunable_2010}.

While this much celebrated interface is a central theme, it is not the only interface currently under study. Charge transfer at the interface between non-superconducting cuprates was also observed, giving rise to interfacial superconductivity~\cite{gozar_high-temperature_2008,logvenov_high-temperature_2009}.

Different effects caused by electron transfer at the interface between a hole-doped Mott insulator and an electron-doped Mott insulator have already been observed in cuprate heterostructures \cite{gozar_high-temperature_2008,jin_anomalous_2011} and the charge redistribution has been the subject of a few models \cite{loktev_model_2008,loktev_superconducting_2011}. There have been  device proposals based on charge redistribution in heterostructures involving Mott insulators \cite{son_heterojunction_2011,millis_electron-hole_2010,hu_proposed_2007}. Also, Mott insulating p-n junctions have been suggested to potentially yield a photovoltaic effect with high energy-efficiency~\cite{manousakis_photovoltaic_2010}. There is some evidence that two overdoped cuprates, one electron-doped, one hole-doped, arranged in a p-n junction geometry, will yield a
high-temperature superconducting effect due to the artificial doping provided by charge redistribution~\cite{hardy_private_2012}. 

Inspired by the experiments on cuprates, here we focus on the electronic charge density distribution and local density of states at the interface between a layered hole-doped Mott insulator and a layered electron-doped Mott insulator. 

Theoretical understanding of such interface problems must first address the question of bulk strongly correlated materials that can exhibit Mott insulating behavior~\cite{imada_metal-insulator_1998}. This requires advanced methods, such as dynamical mean field theory (DMFT)~\cite{georges_dynamical_1996}.  In low dimension, when the self-energy can acquire strong momentum dependence - as happens for example with d-wave high-temperature superconductors - cluster extensions of DMFT such as Cellular Dynamical Mean-Field Theory (CDMFT) or Dynamic Cluster Approximation (DCA) are necessary.~\cite{Hettler:1998,kotliar_cellular_2001,maier_quantum_2005,kotliar_electronic_2006,LTP:2006,SenechalMancini:2011} Potthoff and Nolting \cite{potthoff_surface_1999} initiated research on surfaces and interfaces with DMFT by introducing "inhomogeneous" layered DMFT, that was subsequently applied to various interfaces.~\cite{potthoff_metallic_1999,helmes_kondo_2008,zenia_appearance_2009} 

The case of a Mott/conventional insulator junction was studied in Hartree-Fock theory (HFT) \cite{okamoto_electronic_2004,okamoto_theory_2004} and (DMFT)~\cite{okamoto_spatial_2004,okamoto_interface_2005} by Okamoto and Millis. Similarly, the charge profile of the manganite/cuprate junction \cite{yunoki_electron_2007} and the polar/non-polar Mott insulator junction \cite{lee_electronic_2007} was described with both HFT and DMFT. Magnetic ordering at the interface and the corresponding rich phase diagram can also be explored with simple HFT~\cite{okamoto_electronic_2004,okamoto_theory_2004,yunoki_electron_2007,ueda_electronic_2012-1}. While both HFT and DMFT can be used to obtain an estimate of the charge redistribution, also known as electronic reconstruction, HFT leads to a simple band bending picture while the local density of states as a function of the position obtained from DMFT is non-trivial~\cite{okamoto_spatial_2004,freericks_dynamical_2004,helmes_kondo_2008,zenia_appearance_2009}. 

Here we provide theoretical tools beyond DMFT that are appropriate for layered materials. We take into account both the non-trivial non-local correlations within the individual layers due to strong short-range repulsion using Hubbard model, and the charge redistribution between layers that is dominated by long-range Coulomb repulsion. In-plane quantum fluctuations are included but correlations between the planes are treated at the mean-field (Hartree-Fock) level, neglecting the energy associated to hopping between planes. We call this approach \textit{Dynamical Layer Theory} (DLT). We focus here on the normal state but it will be clear that our approach is easily generalizable to broken-symmetry states. 

The model and method can be found in Sec.~\ref{Model}. The charge density profiles obtained both numerically and analytically, along with density of states profiles can be found in Sec.~\ref{Results}. We focus the discussion in Sec.~\ref{Discussion} on the domain of validity of the method and on the difference between Mott p-n junctions and ordinary semiconductor p-n junctions, concluding with a summary and suggestions for simple extensions of the method.

\section{Model and method}\label{Model}

\begin{figure}[tbp]
\begin{center}
\includegraphics[width=8.5cm]{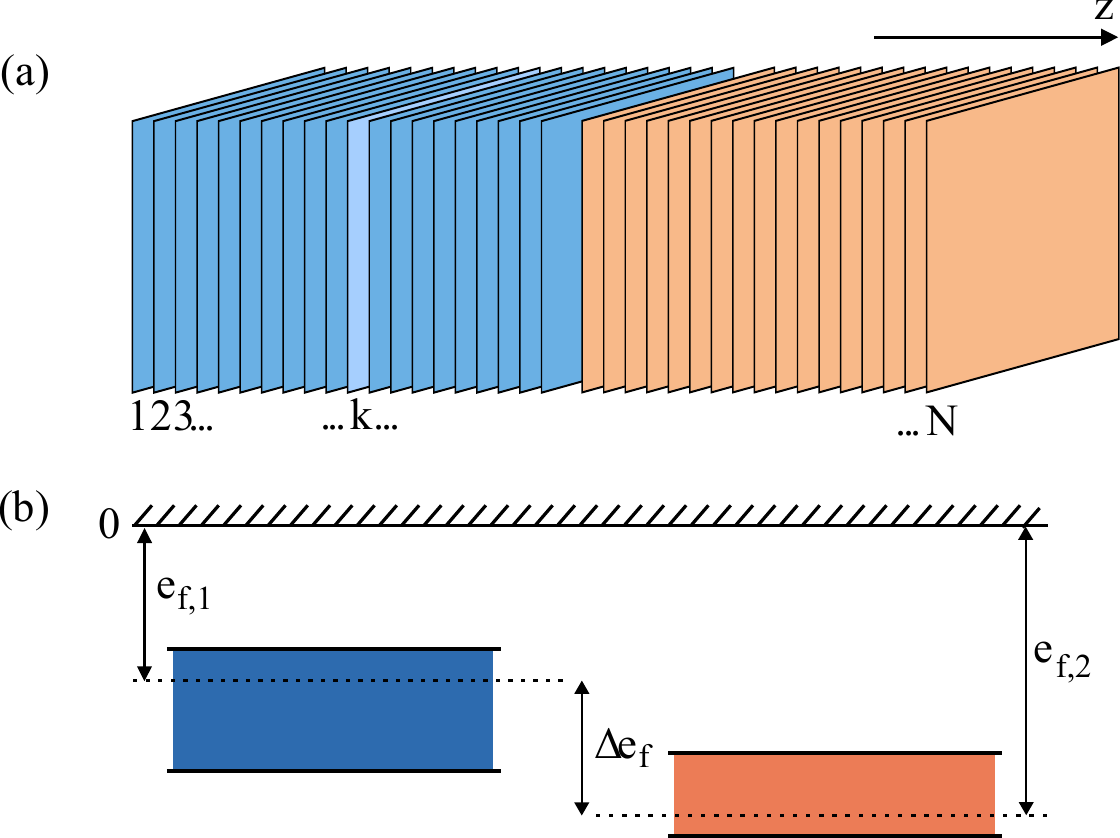}
\end{center}
\caption{(a) Schematic view of the interface. Each plane corresponds to one of the CuO$_{2}$ planes of a cuprate material. The two colors represent 2 different cuprates. (b) Schematic representation of the relevant energies for both materials. On the left is the electron doped PCCO and on the right the hole doped LSCO. Values are not to scale. Due to the Mott nature of these material, there is a gap inside each of the energy bands, but this is omitted here. It is addressed later. Both Fermi energies $e_{f,1}$ and $e_{f,2}$ are negative since we take the vacuum as the zero energy state. $\Delta e_{f}$ is the difference between the Fermi energies.}
\label{fig0:model}
\end{figure}

Although we will consider more general cases, we are motivated by p-n junctions made of a thin film of Pr$_{2-\delta_e}$Ce$_{\delta_e}$CuO$_4$ (PCCO) deposited over a thin film of La$_{2-\delta_h}$Sr$_{\delta_h}$CuO$_4$ (LSCO).~\cite{hardy_private_2012} Fig. \ref{fig0:model} presents a schematic view of this p-n junction with the corresponding definitions of the Fermi energies and of their difference for the bulk systems.

The Hamiltonian for this p-n junction is given by
\begin{equation}
H = H_{\parallel}+H_{\perp}+H_{C}
\label{Htotal}
\end{equation}
where $H_{\perp}$ stands for the hopping between planes, $H_{C}$ for the Coulomb interaction between planes and $H_{\parallel}$ is the Hubbard Hamiltonian of the planes 
\begin{eqnarray}
H_{\parallel}&=&\sum_k  H_{\parallel}^k \nonumber \\
H_{\parallel}^k &=&-\sum_{ij,\sigma}t^k_{ij}c_{i,\sigma }^{k\dag}c^k_{j,\sigma
}+U^k\sum_{i}n^k_{i\uparrow }n^k_{i\downarrow }-\varepsilon ^k_{0}\sum_{i,\sigma
}n^k_{i,\sigma }
\end{eqnarray}%
with $c_{i,\sigma }^{k(\dag)}$ the annihilation (creation) operator at site $i$ in plane $k$ for spin $\sigma$ and $n^k_{i,\sigma}$ the corresponding number operator.  The in-plane hopping matrix $t^k_{ij}$,  Hubbard interaction $U^k$ and site energy $\varepsilon ^k_{0}$ can depend on the material. Here we assume that $t^k_{ij}$ for second neighbor hopping is $t^{\prime}=-0.17t$ \cite{pavarini_band-structure_2001}, $t^{\prime\prime}=0.08t$ for third neighbor hopping and $U=8t$ independent of $k$ while the site energy $\varepsilon ^k_{0}$ is chosen such that we obtain the required bulk filling in plane $k$. Note that in reality, $U$ is somewhat smaller for the electron-doped cuprates when compared with hole-doped ones,~\cite{SenechalHot:2004,Weber:2010} but taking the same value suffices to illustrate the physics. 

Our main approximation is to neglect $H_{\perp}$ in the Hamiltonian Eq.(\ref{Htotal}) compared with $H_{\parallel}$ and $H_{C}$. The large anisotropy in the hopping amplitudes $t_{\perp}\ll t$ of the cuprates motivates this approximation. The physics of charge redistribution will be correctly taken into account when the inequality $\langle H^k_{\perp} \rangle \ll \langle H^k_{C} \rangle$ is satisfied by the final solution. 

The Coulomb interaction coming from the deviation from the bulk charge for the $l$'th plane is given by
\begin{equation}
H_{C}=\frac{1}{2}\sum_{l,m,q_{\parallel}}n_l(q_{\parallel})n_m(-q_{\parallel})V_{l,m}(q_{\parallel}).
\end{equation}
where $n_l(q_{\parallel})$ is the in-plane Fourier transform of the charge-density operator for plane $l$ and $V_{l,m}(q_{\parallel})$ the Coulomb potential. Self-interaction should not be included and it is understood that the uniform background must be subtracted from the charge-density operator for the $q_{\parallel}=0$ contribution to the sum. When charge redistribution occurs on large length scales, the Hartree-Fock approximation is justified for this piece of the Hamiltonian. The Fock term being short range, we assume that it is taken into account by the Hubbard in-plane Hamiltonian. We are then left with 
\begin{eqnarray}
H_{C}&\approx&\frac{1}{2}\sum_{l,m,q_{\parallel}}\langle n_l(q_{\parallel})\rangle n_m(-q_{\parallel})V_{l,m}(q_{\parallel})\nonumber \\
&+&\frac{1}{2}\sum_{l,m,q_{\parallel}}n_l(q_{\parallel})\langle n_m(-q_{\parallel})\rangle V_{l,m}(q_{\parallel})\nonumber \\
&-&\frac{1}{2}\sum_{l,m,q_{\parallel}}\langle n_l(q_{\parallel})\rangle \langle n_m(-q_{\parallel})\rangle V_{l,m}(q_{\parallel})
\end{eqnarray}
where brackets denote thermal and quantum mechanical averages. 

While we could treat cases where the charge distribution breaks lattice translational symmetry in the planes, here we assume uniform solutions. In other words only the $q_{\parallel}=0$ contribution to $\langle n_l(q_{\parallel})\rangle$ survives. Because there are no terms left that involve quantum mechanical operators that couple different planes, the eigenstates are direct products of single-plane states and the grand potential obtained from $\left\langle H\right\rangle$, the ground-state expectation value of $H$, is 
\begin{equation}
\left\langle H\right\rangle - \mu N ={\sum\limits_{k}}\left[E\left( \overline{n}_{k}\right) +%
\frac{\varphi _{k}\left( \overline{n}_{l<k}\right)}{2} \Delta \overline{n}_{k}-\left( \mu
-e_{f,k}\right) )\overline{n}_{k}\right]  \label{ham_tot}
\end{equation}
where $\overline{n}_{k}$ now stands for the expectation value of the $q_{\parallel}=0$ number operator in plane $k$ that we rewrite explicitly as  $\Delta \overline{n}_{k}=\overline{n}_{k}-n_{0,k}$ to emphasize the deviation, mentioned above, with respect to the nominal charge $n_{0,k}$ in the plane. The quantity $E(\overline{n}_{k})$ is defined by $\langle H^k_{\parallel}\rangle$ and the chemical potential $\mu$ is a Lagrange multiplier that guaranties charge neutrality. (Note that if the Fermi energy $e_{f,k}$ (see Fig.(\ref{fig0:model})) is independent of the plane $k$, then the chemical potential is equal to the Fermi energy and all the planes have their nominal bulk filling.) The electrostatic potential energy $\varphi _{k}$ is given by the parallel-plate capacitor formula
\begin{align}
\varphi _{k}\left( \overline{n}_{l<k}\right) & =-\sum\limits_{m
=1}^{k-1}\sum_{l=1}^{m }\frac{\left( e_{C,m }+e_{C,m
+1}\right) }{2}\Delta \overline{n}_{l} \label{Charge_energy}
\end{align}
where we have set the zero of potential on the left-hand side of the junction and defined $e_{C,k} =e^{2}d_{k}/A_{k}\epsilon _{k}$, with $e$ the fundamental charge, $A_{k}$ the area of CuO$_{2}$ in a
unit cell, $d_{k}$ the distance between CuO$_{2}$ planes and $\epsilon
_{k}$ the dielectric constant \cite{chen_electronic_2007}.  

Our problem now reduces to minimizing the functional of the classical variables representing average occupation numbers $\overline{n}_{k}$. The energy $E(\overline{n}_{k})$ is in general a non-linear function of $\overline{n}_{k}$ that can be calculated exactly, or approximately. It contains all the information about the in-plane quantum fluctuations. This is the approach we call Dynamical Layer Theory (DLT). 

In two dimensions, the self-energy near half-filling can acquire non-trivial momentum dependence, even in the normal state. This is why the approximate method that we use to obtain $E(\overline{n}_{k})$ in the layers is CDMFT on a $2\times 2$ plaquette. Single-site DMFT, used in many previous studies, would be inappropriate here. We use an exact diagonalization solver~\cite{Caffarel:1994,Capone:2004,LTP:2006,SenechalBath:2010,SenechalMancini:2011}, then we perform the minimization of the grand potential Eq.~\eqref{ham_tot} to obtain the charge redistribution using both numerical and analytical methods. From this approach, we describe an emerging phenomenon near the interface: a Mott depletion plateau. One can obtain many other properties, including the local density of states profile associated with the charge redistribution and thus the analog of ``band bending'' of the device interface. Many similarities and differences with the classical semiconductor p-n junction are highlighted in the discussion.

\section{Results}\label{Results}

In the first subsection below, we show the $\overline{n}(\varepsilon_0)$ relation and corresponding density of states for a single plane. Numerical and analytical results follow respectively in the two subsequent subsections.

\subsection{Single-plane results}

\begin{figure}[tbp]
\par
\begin{center}
\includegraphics[width=8.5cm]{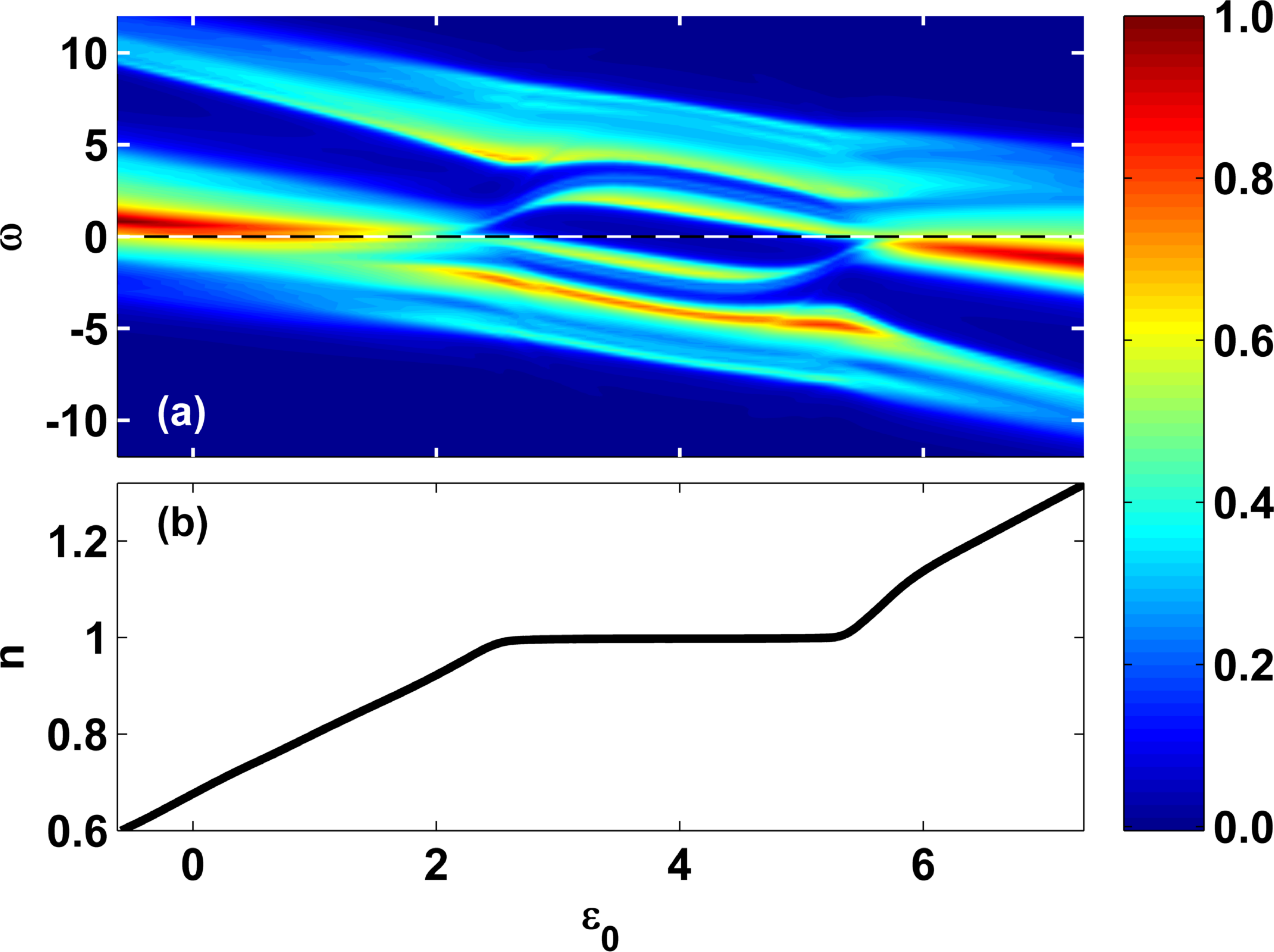}
\end{center}
\caption{(a) Color coded DOS as a function of single-plane chemical potential and
energy. The Fermi level in the DOS is at $\protect\omega=0$ (dashed line).
(b) Corresponding filling ($n$) as a function of single-plane chemical potential
($\protect\mu=\varepsilon_0$ here). The results are for parameters appropriate for LSCO/PCCO, in other words next nearest-neighbour hopping $t^{\prime}=-0.17t$ \cite{pavarini_band-structure_2001}, second nearest-neighbour hopping $t^{\prime\prime}=0.08t$ and $U=8t$. Energy units correspond to $t=1$. We also take $\hbar = 1$.}
\label{fig1:DOS_mu}
\end{figure}

Fig. \ref{fig1:DOS_mu} presents the results for the filling $n$ as a function of $\varepsilon_0$ that plays the role of the single-plane chemical potential $\mu=\partial E/\partial n$ in the case of a single plane. The results are plotted in Fig. \ref{fig1:DOS_mu}. In the
lower panel, the filling $n$ vs $\varepsilon_0$ clearly exhibits
the incompressible Mott phase where $n=1$ as long as $\varepsilon_0$ is in the gap.
The region of zero compressibility, called Mott plateau is
characteristic of a Mott insulator. With the Green function obtained from
the CDMFT calculation we can also compute the density of state (DOS) as a
function of the chemical potential.
It is shown in the top panel of Fig.~\ref{fig1:DOS_mu}. The DOS clearly exhibits the Mott gap present in the "band diagram" for every possible chemical potential. We see that the compressibility ($n^{-2}\partial n/ \partial \mu$) is finite only when the Fermi level ($\omega=0$) crosses a finite DOS. There is an important feature in the DOS around half filling ($n=1$). We note that unlike a traditional band insulator, the DOS is not constant when the chemical potential varies. For example, for electron doping some spectral weight of the lower Hubbard band transfers to the upper Hubbard band at the Mott insulator transition ($\varepsilon_0\sim2.5t,5.5t$).

\subsection{Numerical results for the p-n junction}

\begin{figure}[tbp]
\par
\begin{center}
\includegraphics[width=8.5cm]{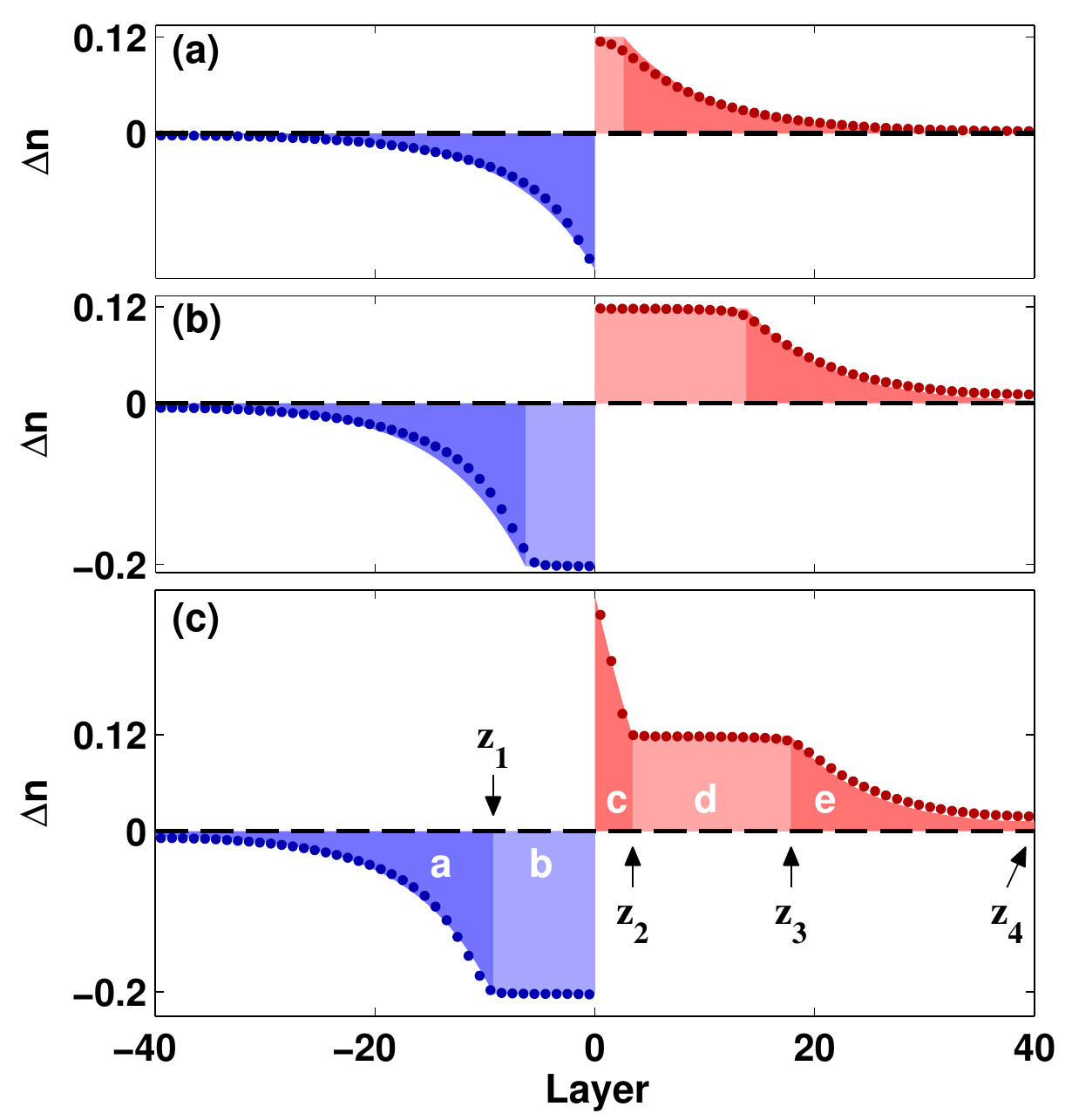}
\end{center}
\caption{Examples of charge redistribution: The charge profile $\Delta n_k$ is plotted as a function of layer number $k$. The parameters are $U=8t$, $e_{S}=0.05t$, $\protect%
n_{0,1}=1.20$, $n_{0,2}=0.88$ and $\Delta e_f=-2t$, $-5.5t$ and $-8t$ for (a), (b) and (c) respectively. These parameters have been chosen to shed light on qualitatively different features of the electronic charge redistribution. Blue is associated to the e-doped Mott insulator
and red is associated to p-doped Mott insulator. The dots are obtained from the
numerical calculation and the shading from the analytically calculated profile
with the same parameters within the continuum approximation. The lighter shading indicate the position of the Mott depletion plateaux.}
\label{fig2}
\end{figure}

A simple numerical minimization algorithm for this Hamiltonian gives
the filling ($\overline{n}_{k}$) for every plane $k$ in the ground state, and thus the
charge redistribution of the junction. Results for an
an electron-doped system on the left side and a hole-doped one on the right side is shown in Fig. \ref{fig2}.
The resulting profiles exhibit some similarities with the classical
semiconductor p-n junction. Nominal doping ($\delta_{0,i}=\left| n_{0,i}-1\right| $), charging energy ($%
e_{C}$) and Fermi level difference ($\Delta e_{f}=e_{f,2}-e_{f,1}$) have been chosen to
reproduce profiles that highlight some similarities and
differences between this Mott p-n junction and semiconductor junctions. The Fermi level is lower for the hole- than for the electron-doped Mott insulator so that charge is transferred from left to right. The Fermi level difference increases in absolute value from panels (a) to (c). 
  
As in semiconductors, there is charge exchange and some
regions can become depleted. Unlike semiconductors, where region $b$ and $d$ of Fig.~\ref{fig2}(c) would extend all the way to the interface and thus be adjacent and form the
well known depletion layer, here there is an additional region $c$ in the
middle. This region behaves like a two dimensional electron gas (2DEG) in the
middle of an Mott insulating depletion layer formed by $b$ and $d$. Outside
the layer formed by $b$, $c$ and $d$, there is Thomas-Fermi screening in
region $a$ and $e$. Semiconductors could in principle be forced into this situation when there exists an appropriate Fermi energy difference, but this does not happen in practice, as far as we know.  

Let us focus on the evolution of the charge redistribution profile caused by the Fermi level
difference ($\Delta e_{f}$) in Fig.~\ref{fig2}. The increase in charge
transfer, caused by an increased Fermi level difference, translates
the reconstructed charge profile, revealing more of the `rigid' curves on each side
without affecting their respective shapes, an effect we will discuss in the following subsection. 

\begin{figure}[tbp]
\vspace{2em}
\par
\begin{center}
\includegraphics[width=8.5cm]{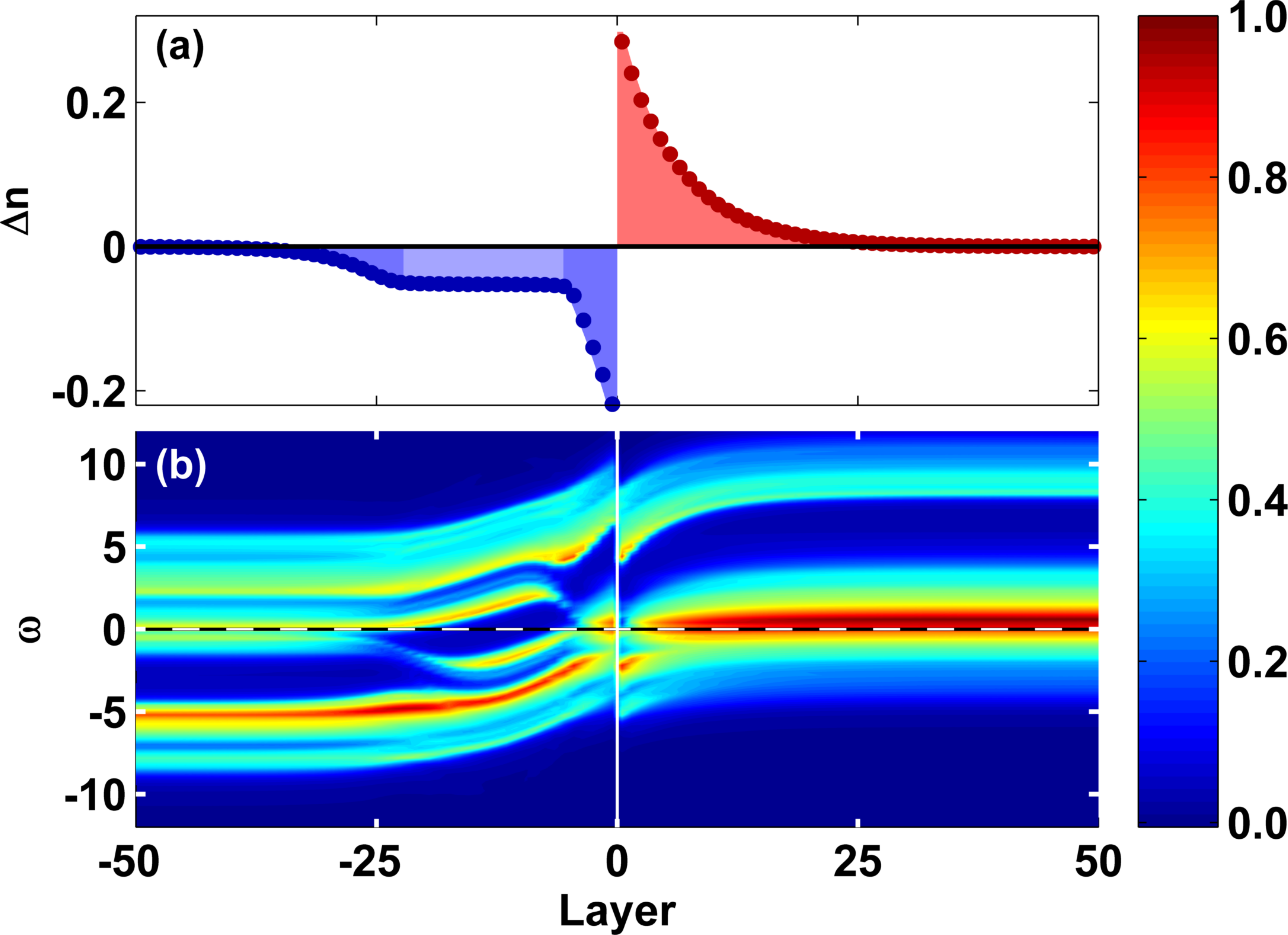}
\end{center}
\caption{(a) Charge redistribution as a function of plane number
(the z-axis) of the structure ($U=8t$, $\Delta e_f =-7t$, $e_{C}=0.1t$).
(b) Color coded DOS in arbitrary units as a function of $\protect%
\omega$ and plane number. The Fermi level in the DOS is at $\protect\omega=0$ (dashed line). We observe an insulating region
several planes wide, where $n=1$.}
\label{fig4}
\end{figure}

A more striking difference with semiconductors arises when one uses the results for the DOS as a function of filling (Top of Fig. \ref{fig1:DOS_mu}), and the filling as a function of the position to compute the local DOS as a function of position. This analog of the semiconductor heterojunction band diagram is illustrated in Fig.~\ref{fig4} for another set of parameters. For this example, nominal doping on each side is chosen in order to obtain only one Mott depletion
plateau on the left part of the junction. Some band bending is observed in
the profile, but there is also a band modulation around the Mott transition
region (on the edge of the Mott depletion plateau). This is caused by the
spectral weight transfer associated to the Mott transition.

Note that there is no need for a doped Mott insulator on the right-hand side of Fig. %
\ref{fig4} to obtain a Mott depletion plateau on the left. The right-hand part of
the structure could easily be a semiconductor or an insulator with appropriate
relative Fermi level. As long as there is sufficient charge transfer, there
will be this plateau. In other words, the resulting charge profile on
one side is not only rigid, but also its shape is independent of the
material on the other side of the junction that attracts (or repels) the charges.




\subsection{Analytical results for the p-n junction}

For a large number of CuO$_{2}$ planes, a numerical approach is needed to obtain the filling from the minimisation of Eq.~\eqref{ham_tot}. Nevertheless, the
profile can be estimated analytically by taking the continuum limit $\sum_a^b \rightarrow \frac{1}{d}\int_a^b dz$. In
this limit, the minimization procedure consists in finding the charge
distribution $n\left( z\right)$ where the functional derivative
of the ground state energy Eq.~\eqref{ham_tot} with respect to the charge distribution $n\left(z\right)$ vanishes. This leads to a Thomas-Fermi equation 
\begin{equation}
\Delta\mu\left( z\right) =\left( \mu - e_{f,z} + \frac{1}{2}\Delta\varphi_{tot}\right) - \varphi\left(z\right)  \label{delta_mu1}
\end{equation}%
where we defined $\Delta\mu\left( z\right) \equiv \delta E\left(n(z)\right)/\delta n(z)$ and $\Delta\varphi_{tot}$ as the total electrostatic potential-energy difference between the right-hand side and the left-hand side of the junction. This term arises from the functional derivative of the electrostatic potential energy
\begin{equation}
\frac{\delta \left( \int dz'\varphi(z')\Delta n(z')\right)}{\delta n(z)}=2\varphi (z) - \Delta\varphi_{tot}.
\end{equation} 
In the limit where both sides of the junction are semi-infinite, we have
\begin{align}
\Delta\varphi_{tot}  &  =e_{f,2}-e_{f,1}\\
\mu &  =\left(  e_{f,2}+e_{f,1}\right)  /2. 
\end{align}

The second derivative of the Thomas-Fermi equation Eq.~\eqref{delta_mu1} gives us an equivalent of the Poisson equation $d^2~ \partial ^{2}\Delta \mu \left( z\right) / \partial z^{2} = - e_{C} \Delta n\left( \Delta \mu \left( z\right) \right)$. The main source of simplification comes from the fact that the relation between $\Delta n$
and $ \Delta \mu \left( z\right)$, which can be
obtained from to the lower panel of Fig. \ref{fig1:DOS_mu}, can be
approximated by a piecewise continuous function that is either linear or constant in the vicinity of half-filling. 

The differential equation can then be solved separately in different segments with boundary conditions between the segments given by the
continuity equation for the potential and the electric displacement. The
analytical solution for the case illustrated in Fig. \ref{fig2}(c), for example, appears as the shaded area.
In regions $c$, $d$ and $e$ on the right-hand side, the charge profile is a piecewise continuous function given by
\begin{eqnarray}
\Delta n_{c}\left( z\right)  &=&\delta_{0,2}\left[ e^{\frac{-\left(
z-z_{2}\right)}{\lambda _{TF}} }+\frac{\Delta z}{\lambda _{TF}}\sinh \left( \frac{-\left(
z-z_{2}\right)}{\lambda _{TF}}
\right) \right]    \nonumber \\
\Delta n_{d}\left( z\right) &=&\delta_{0,2}  \nonumber\\
\Delta n_{e}\left( z\right)  &=&\delta_{0,2}\left[ e^{\frac{-\left(
z-z_{3}\right)}{\lambda _{TF}} }+g\left( z\right) \right]­.   \label{profile3}
\end{eqnarray}%
In the depletion region $d$, the electron density difference $\Delta n_{d}$ is equal to the nominal hole doping of the material $\delta_{h}$. The term $g\left( z\right) =\exp \left[ \left( z-\left(
2z_{4}-z_{3}\right) \right) /\lambda _{TF}\right] $ is negligible for
structures that are thick enough ($z_{4}\rightarrow \infty $). $2z_{4}-z_{3}$
is the mirror position of $z_{3}$ with respect to the right end of the structure.
The difference between $z_{3}$ and $z_{2}$, the length of the Mott depletion
plateau, is given by
\begin{equation}
\Delta z=\lambda _{TF}(\sqrt{1+2\left\vert E_G/\delta \mu
_{2}\right\vert }-1)  \label{delta_z}
\end{equation}%
where $E_G$ is the Mott energy gap, $\delta \mu _{i} \equiv \delta_{0,i}\partial \mu/\partial n $ the chemical potential difference between the nominal filling and half-filling, while $\lambda _{TF}=d/\sqrt{e_{C}~\partial n/\partial \mu}$ is the Thomas Fermi length-scale for
the charge profile, all in the approximation $\left. \partial n/\partial \mu \right\vert
_{n\neq 1}$ constant outside the Mott plateau. The result for the size of the Mott plateau $\Delta z$, comes physically from the fact that there is an electric field that changes linearly with distance inside the plateau leading to an electrostatic potential difference that grows quadratically with distance. When the electrostatic potential energy difference becomes equal to the Mott energy gap $E_G$, the plateau ends.

The function $g\left( z\right)$ in Eq.~\eqref{profile3} allows the electric field to vanish at the surface of the junction. Here, we want to stress the idea that, except for $%
g\left( z\right) $, the shape of the piecewise continuous function describing the electronic density profile is rigid and
does not deform with charge transfer: A larger $\Delta e_{f}$ only reveals
more of the curve (Eqs.~\eqref{profile3}). The only remaining unknowns for the electronic density profile is the position $z_{2}$
(and hence $z_{3}$ that just follows from $\Delta z$) and $z_{1}$ relative to the interface. These can be found from the continuity of the electric
displacement at the interface and from $\Delta\mu(0^-)-\Delta\mu(0^+)=\Delta e_{f}=e_{f,2}-e_{f,1}$ that follows from the Thomas-Fermi Eq.~\eqref{delta_mu1}. Note that the energy
scale $e_{C}$ is very difficult to obtain from first principles. It will vary considerably depending on material. Its value changes considerably the length-scale of the charge profile.

\section{Discussion}\label{Discussion}

The Mott depletion plateau obtained here is consistent with the result obtained by Oka and Nagaosa~\cite{oka_interfaces_2005} even though these authors used a completely different technique, namely the one-dimensional density-matrix renormalization group. In other words, our method, which neglects hopping energy between planes, predicts the same incompressible state as a full one dimensional treatment. This plateau is also present in some HFT \cite{yunoki_electron_2007} and DMFT \cite{ishida_embedding_2009} studies.

We can verify the validity of our approach (DLT) in the vicinity of the Mott plateau by comparing the hopping energy $t_{\perp}$ with the potential energy difference between two adjacent planes in the plateau $\sim d~\partial\Delta\mu (z)/\partial z$. We can estimate the latter from the smallest value of the derivative, namely at the beginning of the Mott plateau. Since the profile just before the plateau is exponential, we find that
\begin{equation}
t_{\perp} \ll \delta \mu_i (d_i / \lambda_{TF,i}) = \delta_{0,i} e_{C,i} (\lambda_{TF,i} / d_i) \label{condition_approx}
\end{equation}%
must be satisfied for both sides $i$ of the p-n junction. Since $\lambda_{TF,i} \geqslant d_i$, $e_{C,i} \gtrsim t$ and $t \gg t_{\perp}$ in cuprates, this condition will hold even for low dopings $\delta_{0,i}$. However, we see that for nominal dopings $\delta_{0,i}$ that are too small, this condition will not hold. Indeed, we see from the expression for the size of the Mott plateau, Eq.{\eqref{delta_z}, that small nominal dopings $\delta_{0,i}$ imply a very large Mott plateau, consistent with small charge transfer and hence small long-range Coulomb interaction. Note that in the examples used in this work, the relative dielectric constant is much larger than that which is usually quoted for the cuprates ($\approx 10$ to $20$). This leads to much larger values of the Thomas-Fermi screening length than the value $\lambda_{TF,i} \approx d_i$ expected in the cuprates.~\cite{logvenov_high-temperature_2009}

There is an important parallel that can be drawn between this Mott p-n
junction and the classical semiconductor p-n junction. The charge profile at
the interface in middle panel of Fig.~\ref{fig2} corresponds to the depletion
approximation in semiconductors. There are however three major differences.

1. The Mott gap can be crossed at the interface and a two-dimensional gas
of charge carrier can appear in the middle of a depletion layer like region $%
c$ in Fig.~\ref{fig2}. Using the equation for the size of the depletion layer Eq.~\eqref{delta_z} but for semiconductors,
the doping energy $\delta \mu_2$ is so small compared with the gap energy $%
E_G$ that the length of the depletion plateau becomes very large.

2. In a Mott p-n junction, the length-scale for the charge redistribution 
is much smaller because the charging energy $e_{C}$ is generally much higher. 

3. The most important difference is that in addition to band bending, there is also band modulation around the Mott depletion plateau region in contrast with rigid band bending in semiconductors. This might lead to a broader spectrum of light absorption and emission.

We emphasize that the whole analysis has treated $e_{C}$ and $\Delta
e_{f}$ as free parameters. For PCCO and LSCO, the parameter $e_{C}$
should be around ten times larger than what we have used. The resulting
profile would be ten times shorter that what we showed here. We do not know
precisely the value of $\Delta e_{f}$, but it should be of the order of the
kinetic energy $\sim t$. We need a $\Delta e_{f}$ larger
than the Mott gap $E_G$ to obtain a two-dimensional gas of charge carrier. This might be achieved through
the application of a large potential difference. Overall, our results are not limited to cuprates but apply to all layered doped Mott insulators. 

\section{Conclusion}

In summary, we developed a very simple and intuitive method, Dynamical Layer Theory, to account for Hubbard model correlations and calculate the charge profile, local DOS profile and other observables of layered correlated systems. The main approximation is to neglect interlayer hopping compared with the charging energy treated in the Hartree-Fock approximation. Once the solution is found as a function of chemical potential in the various planes, we are left with a minimization problem to find the filling of each plane. Here we found the solution of the Hubbard model within the planes using plaquette CDMFT with an exact diagonalization solver. We neglected broken symmetries and finite temperature effects, but they can all easily be included within our approach. This opens, for example, the possibility to treat d-wave superconductivity accurately in such structures. 

Both numerical an analytic calculations predict an incompressible Mott depletion layer analog to the semiconductor depletion layer, and the possibility to obtain quasi two-dimensional charge carriers of opposite signs on either side of the interface. The calculated density of states near the Mott depletion plateau deviates from expectations from simple band bending ideas and should have interesting consequences for the absorption and emission spectrum of this device.



We are indebted to Patrick Fournier, Maxime Dion and Guillaume Hardy for sharing their experimental results and for numerous discussions. Work at Sherbrooke was partially supported by a Vanier Scholarship from NSERC (M.C.) and by the Tier I Canada Research Chair Program and NSERC (A.-M.S.T.). Computing resources were provided by CFI, MELS, Calcul Qu\'ebec and Compute Canada.

\bibliographystyle{./h-physrev3_v4}
\bibliography{./ref_papier3}

\end{document}